\documentclass[epj]{svjour}
\usepackage{graphics}
\usepackage{amsmath}
\begin{document}
\title{GIARPS@TNG}
\subtitle{GIANO--B \& HARPS--N together for a wider wavelength range spectroscopy}
\author{R. Claudi \inst{1} \thanks{\emph{Present address:} INAF -- Oss. Astronomico di Padova, Italy} \and S. Benatti\inst{1} \and  I. Carleo \inst{1, 2} \and A. Ghedina \inst{3} \and   J. Guerra \inst{3} \and  G. Micela \inst{4}  \and E. Molinari \inst{3, 5} \and  E. Oliva \inst{6} \and M. Rainer \inst{7} \and  A. Tozzi \inst{6} \and  C. Baffa \inst{6} \and  A. Baruffolo \inst{1} \and  N. Buchschacher \inst{8} \and  Cecconi M. \inst{3} \and  R. Cosentino \inst{3} \and  D. Fantinel \inst{1} \and L. Fini \inst{6} \and  F. Ghinassi \inst{3} \and E. Giani \inst{6}  \and E. Gonzalez \inst{4}  \and M. Gonzalez \inst{3}  \and R. Gratton \inst{1} \and  A. Harutyunyan \inst{3} \and  N. Hernandez \inst{3} \and  M. Lodi \inst{3} \and  L. Malavolta \inst{2} \and J. Maldonado\inst{4} \and L. Origlia \inst{9} \and N. Sanna \inst{6} \and J. Sanjuan \inst{3} \and S. Scuderi \inst{10} \and U. Seemann \inst{11} \and A. Sozzetti \inst{12} \and H. Perez Ventura \inst{3} \and M. Hernandez Diaz \inst{3} \and  A. Galli \inst{3} \and C. Gonzalez \inst{3} \and L. Riverol \inst{3} \and  C. Riverol \inst{3}
}                     
\offprints{R. Claudi (riccardo.claudi@inaf.it)}          
\institute{INAF Astronomical Observatory of Padova, Italy (riccardo.claudi@inaf.it) \and Dep. of Physics and Astronomy, University of Padova, Italy \and INAF, Fundacion Galileo Galilei, Spain  \and INAF Astrophysical Observatory of Palermo, Italy  \and  INAF-IASF Milano, Italy  \and  INAF Astrophysical Observatory of Arcetri, Italy  \and  INAF Astronomical Observatory of Brera, Merate, Italy  \and  University of Geneva, Switzerland  \and  INAF Astronomical Observatory of Bologna, Italy  \and  INAF Astrophysical Observatory of Catania, Italy  \and  Georg-August Universit\"at G\"ottingen, Institut f\"ur Astrophysik, G\"ottingen, Germany \and  Astrophysical Observatory of Torino, Pino Torinese, Italy}
%
\date{Received: date / Revised version: date}
%
\abstract{
Since 2012, thanks to the installation of the high resolution echelle spectrograph in the optical range HARPS-N, the Italian telescope TNG (La Palma) became one of the key facilities for the study of the extrasolar planets. In 2014 TNG also offered GIANO to the scientific community, providing a near-infrared (NIR) cross-dispersed echelle spectroscopy covering $0.97-2.45\ \mu$m at a resolution of 50,000. GIANO, although designed for direct light-feed from the telescope at the Nasmyth--B focus, was provisionally mounted on the rotating building and connected via fibers to only available interface at the Nasmyth--A focal plane. The synergy between these two instruments is particularly appealing for a wide range of science cases, especially for the search of exoplanets around young and active stars and the characterisation of their atmosphere. Through the funding scheme "WOW" (a Way to Others Worlds), the Italian National Institute for Astrophysics (INAF) proposed to position GIANO at the focal station for which it was originally designed and the simultaneous use of these spectrographs with the aim to achieve high-resolution spectroscopy in a wide wavelength range ($0.383 - 2.45\ \mu$m) obtained in a single exposure, giving rise to the project called GIARPS (GIANO-B \& HARPS-N). 
Because of its characteristics GIARPS can be considered the first and unique worldwide instrument providing not only high resolution in a large wavelength band, but also a high precision radial velocity measurement both in the visible and in the NIR arm, since in the next future GIANO-B will be equipped with gas absorption cells.
\keywords{Instrumentation; Spectroscopy; Extrasolar Planets}
} 
\maketitle
\section{Introduction}
The discovery of a hot Jupiter orbiting 51Peg\cite{mayorequeloz1995} triggered the quest for extrasolar planets using the radial velocity technique, by which the presence of a planetary companion is inferred by the wobble it induces on the parent star. In the present, the search for more exotic stellar hosts and the race for the lightest planets point towards M dwarfs. These stars, which are the most abundant in the Universe, are also the smallest ones. Since the RV variation induced by a planet on a star scales with M$_{\star}^{-2/3}$, the amplitude of the effect induced on an M star is significantly larger. As an example, a planet of an identical mass at the same distance from the stellar host produces a RV variation with an amplitude $\sim 3$ times larger on an M5 star than on a G2 star. The drawback is that since they are much colder, M dwarfs are much fainter in optical wavelengths. The RV surveys of light-mass stars points then towards the exploration of a new wavelength domain, the near-infrared (NIR), where the luminosity of these objects peaks. This type of stars can show surface inhomogeneities like stellar spots, being young and active, which can mimic or hide Doppler signal due to a planet. Observing in the NIR, as opposed to VIS, the contrast between these surface inhomogeneities and the stellar disk is strongly reduced,  helping to discriminate between colored signal (activities, pulsations etc.) and planetary signal. This highlights another advantage of measuring radial velocities in NIR.

In this framework GIARPS (GIAno \& haRPS-n) \cite{claudietal2016}, the new common feeding for both the high resolution spectrographs, HARPS-N in the visible and GIANO in the NIR, represents a good chance to investigate this class of objects in the next future. GIARPS will allow to have the two instruments on the same focal station of the Telescopio Nazionale Galileo (TNG) working simultaneously. This allows to have such a unique facility in the north hemisphere on duty at the TNG. 

\section{Science case}
The GIARPS science case is very broad, given the versatility of such an instrument and the large wavelength range. A number of outstanding science cases encompassing mainly extra--solar planet science starting from rocky planet search and hot Jupiters atmosphere characterization can be considered. In particular:
\begin{description}
\item[{\bf Observation of hot planet atmospheres.}] Transiting extrasolar planets present a unique opportunity to study their atmospheres. Wavelength dependent variations in the height at which the planet becomes opaque to tangential rays will result in wavelength-dependent changes in the ratio of spectra taken in and out of the transit. Several groups (\cite{vidalmadjaretal2003}; \cite{vidalmadjaretal2004}; \cite{tinettietal2007}, \cite{swainetal2008}; \cite{swainetal2009}; \cite{beaulieuetal2009}, \cite{tinettietal2010}) have pursued transmission spectroscopy and secondary occultation measurements detecting absorption signatures of several atoms (Hydrogen, Oxigen and Carbon in the visible and UV) and molecular signatures (in near and mid IR) from water, methane, Carbon monoxide and dioxide. While most of these results have been obtained by space-based observations, several efforts to obtain ground based observations have been performed in the recent years, resulting in ground-based detection of transmission features (\cite{redfieldetal2008}, \cite{snellenetal2008}, \cite{singetal2009a}) and optical and near IR eclipse measurements (\cite{singandlopez2009}, \cite{demooijandsnellen2009}, \cite{crolletal2010}). Ground-based transmission spectroscopy could also be very fruitful in the near IR, where high spectral resolution can resolve molecular absorption bands and can be used to combine the signal of tens to hundreds of individual molecular absorption lines (from e.g. water, CO, or methane). In this framework GIARPS with its resolution and wavelength range extending from visible to near IR represents a good opportunity to obtain spectroscopic information on hot small mass companion atmospheres. The molecules that we will mainly be targeting with GIARPS are O$_2$, H$_2$O, CO$_2$ and CH$_4$, which are all molecules with large broad absorption lines in the visual to near-infrared part of the Earth transmission spectrum (\cite{palleetal2009}). In addition, during secondary eclipse, the planet passes behind the star and we can measure the planet's radiation directly by subtracting the photometric measurement during an eclipse from the measurement before or after the eclipse. Of about 1,000 transiting exoplanets, only a small fraction are bright enough (V$<12$) to be suitable for atmospheric studies, but when Cheops and TESS will be launched in $\sim 2017$ many more systems around bright star will be found.

\item[{\bf Search for Giant planets around young nearby stars.}] Giant planets (GPs) play a crucial role for (exo)-planetary systems as they shape their final architecture because they strongly impact the dynamics and the fate of lighter bodies in the system. They represent a significant part of the planets detected so far. However, we are far from having a complete knowledge of their occurrence, diversity and properties as GP exploration with RV or transit techniques is limited today to typically 5 au from the stars (and even closer, 1 au, for transiting planets). Their outer population will remain out of reach of deep imagers until the ELT era. Young stars (a few Myr to a few hundreds Myr) are the only sources that could, in the near future, allow complete (mass-period) exploration of the giant planet population, using both current and next generation of planet imagers and high precision spectrographs. In fact, while young GP have not completely cooled out and are still hot enough to be detected by direct imaging, they are also within the reach of high resolution spectrographs, given that good enough precision can be reached despite the high activity levels of the host stars. Given their ages, young stars are also unique laboratories to study on -- going formation processes and the building--up of planetary systems and get direct information on the associated timescales (e.g. migration of hot-Jupiters). In addition, young systems give the possibility to test and calibrate evolutionary models, and in particular the controversial mass-luminosities predictions at young ages: ''hot- start'' and ''cold-start'' models, which differ by the respective amounts of energy lost/available to heat the planet at early stages (\cite{fortneyetal2008}). In this context, coupling imaging to RV data offers a unique opportunity to do so, by directly estimating/constraining the dynamical masses of imaged planets, as done recently in the case of $\beta$\ Pic\ b (\cite{lagrangeetal2012}). Up to now, young stars have been mostly avoided in RV surveys as they are usually active, hence 1) more rapid rotators than their older counterparts (hence not suited for masking techniques), and 2) their stellar jitters are higher than those of their Main Sequence counterparts, which a priori alters planet detectability. Several studies (see e.g., \cite{huelamoetal2008}\cite{baileyetal2012}) have demonstrated that the stellar jitter is significantly lower in the IR, even for the most active stars. Also, it has been demonstrated that high resolution IR spectrographs as VLT/CRIRES can reach precisions down to 5 m/s (see e.g., \cite{seifhartandkaufl2008}, \cite{Beanetal2010}), allowing to detect close companions well within the planetary regime. The simultaneous use of high resolution and wide wavelength range spectrographs like GIARPS, offer the unique opportunity to combine both techniques to constrain the giant planet population at all periods and test theories of planetary formation and evolution.

\item[{\bf Search for hot planets around cool stars.}] The study of M-type stars is gaining momentum as an alternative ÔÕfast trackÕÕ method to discover and possibly characterize hot and temperate rocky exoplanets. M--type stars are the most abundant type of stars in our Galaxy, and therefore obtaining statistics of planet occurrence and architecture around these stars is of great importance for understanding the physics of planet formation and evolution and its dependence on stellar host mass. Planet searches around M--type stars (with masses in the range of $0.1 \div 0.6$\ M$_\odot$) have the main advantage of the larger RV signal, the smaller star-planet contrast and the shorter orbital period of the Habitable Zone (HZ). This has been exploited to find some of the low-mass Super-Earth exoplanets known so far both with RV (\cite{mayoretal2009}, \cite{angladaescudeetal2012}) and transits (\cite{charbonneauetal2009}; \cite{bonfilsetal2013}); although the current number of detections is still low compared with solar analogues. In spite of that, recent studies from ground--based RV programs carried out with state--of--the--art facilities (HARPS) indicate that Super--Earths with 6--10 Earth masses within the HZ of low- mass stars appear ubiquitous (\cite{bonfilsetal2013}).  Recent analyses of Kepler data have only further corroborated this evidence (\cite{dressingandcharbonneau2013}; \cite{kopparapu2013}). For the solar-neighbourhood sample, in particular, the abundance of planets as a function of mass and orbital distance is very loosely constrained. In addition, all the results obtained from RV surveys are only valid for M-type stars of spectral types earlier than M2 or M3. The faintness of the targets and the intrinsic stellar jitter have traditionally limited the investigation of even lower mass stars. Near -- IR, high--precision ($< 10$\ m/s) RV observations can be more efficiently used to detect low--mass (potentially habitable) planets around stars with spectral types later than about M3. RV measured in the near--IR are also less susceptible to stellar radial--velocity noise.
\end{description}
Not only exoplanets but also other science cases like  young stars and proto--planetary disks, cool stars and stellar populations, moving minor bodies in the solar system, bursting young stellar objects, cataclysmic variables and X-ray binary transients in our Galaxy, supernovae up to gamma--ray bursts in the very distant and young Universe, can take advantage of the unicity of this facility both in terms of contemporaneous wide wavelength range and high resolution spectroscopy.

\section{The Visible arm: HARPS--N}
\begin{figure}
\centering
\resizebox{0.75\textwidth}{!}{
 \includegraphics{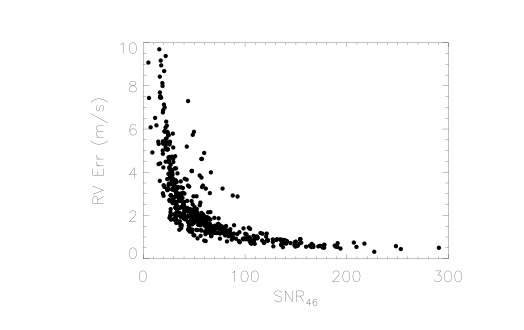}
}
\caption{Typical internal errors of the HARPS--N radial velocity measurement as function of the SNR at the central order of the spectra.}
\label{fig:harpRV}       
\end{figure}

Since April 2012 the high resolution echelle spectrograph HARPS--N \cite{cosentinoetal2012} is part of the equipment of the Telescopio Nazionale Galileo (TNG), after an agreement between the Italian National Institute of Astrophysics (INAF) and the HARPS--N Consortium\footnote{Composed by Geneva Observatory (CH), INAF-TNG (Italy), CfA and Harvard University (USA), University of St. Andrews, Edinburgh, and Belfast (UK).}. HARPS--N is the Northern counterpart of HARPS \cite{mayoretal2003}, mounted at the ESO 3.6 m telescope in La Silla (Chile). Both of them are characterized by an extreme instrumental stability that allows radial velocities measurement with the highest accuracy now available, even below 1 m/s. Figure \ref{fig:harpRV} shows the internal errors of the HARPS--N radial velocity measurement as function of the SNR at 550 nm, for the spectra collected in the framework of the GAPS project (Global Architecture of Planetary Systems, \cite{covinoetal2013}). For high values of SNR the lower envelope of the distribution is less than 1 m/s.

HARPS--N is mounted at the Nasmyth B focus of TNG, and it covers the spectral domain from 390 to 690 nm with a mean resolution of 115,000. The instrument is composed by two main parts: the spectrograph, which is located in a separated room on the ground floor of the telescope inside a vacuum vessel to minimize the vibrations and to ensure the required stability of temperature and pressure, and the Front End Unit (FEU), which is mounted on the Nasmyth B fork. These two modules are connected through an optical fiber link. The FEU is the first part of the instrument, where the incoming light from the telescope (or the calibration lamps) is conditioned and collimated into the fibers. The star is maintained inside the fiber thanks to a tip-tilt mirror acting together with the auto guider system. The octagonal fiber link increases the light scrambling effect and guarantees a very high precision in radial velocity measurement, since it minimizes the spectrograph illumination changes due to the positioning error of the star in the fiber entrance. The two fibers have 1 arcsec aperture each: one is dedicated to the scientific object, while the second one is used for reference (background sky or a Fabry-Perot interferometer). The fiber entrance is re-imaged by the spectrograph optics onto a 4k$\times$4k CCD, where echelle spectra of 69 orders are formed for each fiber. The spectrograph is mounted on a nickel plated stainless steel mount and contains no moving parts.

The HARPS--N integrated pipeline \cite{Sosnowskaetal2012} provides to the observer a complete reduced dataset only 25 seconds after the end of the exposure. The data reduction pipeline takes into account the data images (calibration, bias, dark and scientific), performs quality control and executes a complete data reduction. The radial velocity measurement is performed through the application of the cross-correlation function (CCF) to the acquired spectrum of a specific mask that depicts the typical features of stars with different spectral types (see \cite{pepeetal2002}).
The output is a set of data including the reduced frames, wavelength-calibrated spectra and the results of the CCF method (FWHM, radial velocities, bisector velocity span, etc).

\section{The NIR arm GIANO}
At the beginning of 2015, TNG offered for the first time to the scientific community GIANO (\cite{olivaetal2006}), its new near infrared (NIR) high resolution echelle spectrograph. After the commissioning and science verification observing runs in 2013 and 2014, GIANO demonstrated its capability to fulfill the required performances, reaching for instance a satisfying accuracy for the radial velocity measurements ($\sim 10$m/s, \cite{carleoetal2016}). A single exposure with GIANO produces a spectrum ranging from Y to K band ($0.95 - 2.45 \mu$m) with a resolution of 50,000 \cite{origliaetal2014}. 
GIANO is currently mounted at the Nasmyth A focus of  the TNG and fed by IR-transmitting ZBLAN fibers. The instrument is composed by two main modules: the cryogenically cooled spectrograph and the warm preslit and interface system. The preslit system also includes a fiber mechanical agitator and rackmount with the electronics and the calibration lamps. An additional rack--mount contains the detector warm electronics and controls.
The GIANO spectrometer is mounted on a rigid aluminum bench, thermally connected to a LN2 tank. Following the light path of the instrument from the entrance window, the spectrometer includes a flat window, a cold stop, a filters wheel, a slit, the spectrometer optics (7 mirrors, 3 prisms, 1 grating) and a 2k$^2$ HgCdTe detector array. 
\begin{figure}
\centering
\resizebox{0.7\textwidth}{!}{
 \includegraphics{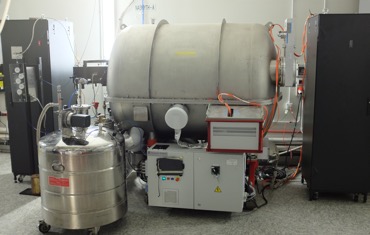}
}
\caption{GIANO vacuum vessel, fiber agitator, LN2 tank and the electronics in the Nasmyth A of TNG \cite{claudietal2016}}
\label{fig:oldgiano}       
\end{figure}
All of these elements are included inside a vacuum chamber (see Figure \ref{fig:oldgiano}) which is permanently connected to all the sub-systems (pipelines, valves, pumps, sensors, PLC), necessary to create, maintain, monitor and control the vacuum and the cryogenic status of the spectrometer. All the operations are performed and supervised by the PLC, controlled through a dedicated panel. 
The GIANO preslit boxes are optically connected by a couple of fibers optics, used to simultaneously observe the target and the sky. The core diameter of each fiber is 85 $\mu$m (1 arcsec on sky), the distance between the two fiber centers is 250 $\mu$m (3 arcsec on sky). A third fiber is used for calibration, and can be illuminated either by a halogen lamp for flat--field or by an U--Ne lamp for wavelength calibration. The introduction of the fibers, due to constraints imposed on the telescope interfacing during the pre--commissioning phase, has significantly reduced the end--to--end efficiency of the instrument. It also introduced a non repeatable spectral modulation that cannot be corrected by flat-fielding, limiting the signal to noise ratio achievable in the spectra, regardless of the brightness of the star and integration time. For this reason GIANO uses a mechanical agitator to decrease the effect of fiber modal noise. This mechanism works quite well for diffuse sources like the calibration lamps, but for observation of scientific targets, the modal noise is amplified by effects related to the non-uniform illumination of the fiber (which also depends on the seeing conditions and on the tracking/guiding performances of the telescope). In stellar spectra acquired in non optimal observing conditions the residual modal noise can be as high as a few percent, especially at longer wavelengths (K band), thus limiting the overall signal-to-noise ratio.

\begin{figure}
\centering
\resizebox{0.35\textwidth}{!}{
 \includegraphics{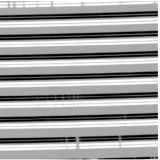}
}
\caption{Science 2D -- spectrum obtained with the sum of each pair of A--B spectra.}
\label{fig:framegiano}       
\end{figure}

The GIANO echellogram has a fixed format and includes about 50 orders, covering the $0.95-2.45 \mu$m wavelength range. It has a full spectral coverage up to 1.8 $\mu$m, while at the longest wavelengths the spectral coverage is about 75\%. Due to an image slicer, each 2D frame contains four tracks per order. 
Observations of science targets are generally performed by nodding-on-fiber, i.e., target and sky are taken in pairs and alternatively acquired on fiber A and B, respectively. From each pair of exposure an ``A--B 2D--spectrum" is computed, then extracted and summed together for an optimal subtraction of the detector noise and background. The positive (on A fiber) and negative (on B fiber) spectra of the target star are shown in Figure \ref{fig:framegiano}.

\section{GIARPS}
\begin{figure}
\centering
\resizebox{0.7\textwidth}{!}{
 \includegraphics{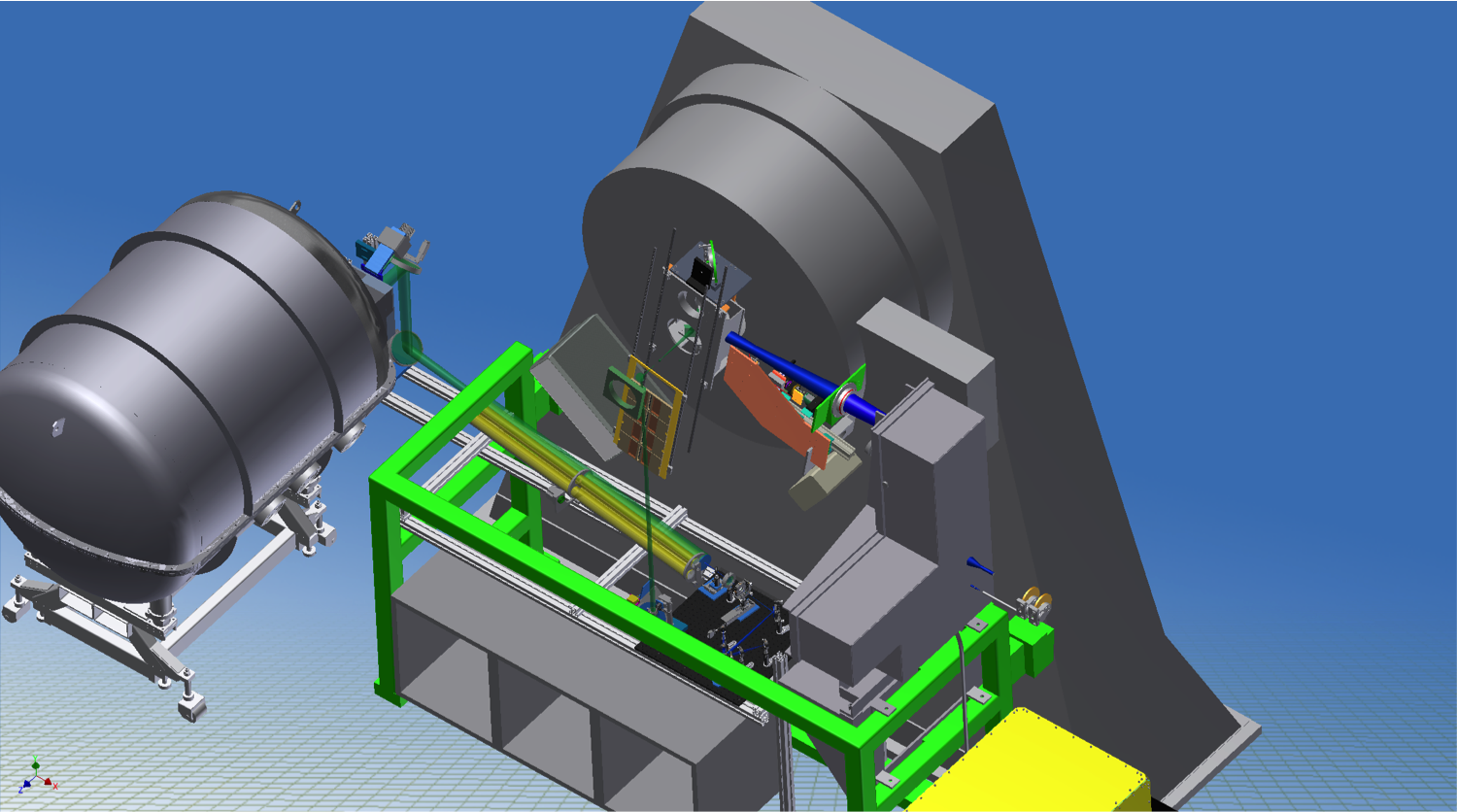}
}
\caption{Mechanical solution for the housing of GIANO (left side) in Nasmyth B. The new preslit is located inside the SARG gage (green structure).}
\label{fig:structure}       
\end{figure}

GIARPS is the new configuration of GIANO and HARPS--N that will allow the simultaneous use of the two spectrographs, exploiting therefore a wide wavelength range ($0.390 - 2.45 \mu$m) with high resolution (115,000 in the visible and 50,000 in the NIR) obtained in a single exposure. The two instruments will be also able to work separately, so with GIARPS the TNG will provide three different high resolution spectroscopy observing modes: a) HARPS--N only (maintaining the current optical configuration with the already existing mirror); b) GIANO only; c) both GIANO and HARPS--N splitting the light with a dichroic. 
The simultaneous use of the two spectrographs can be obtained by moving GIANO from Nasmyth A to Nasmyth B focus of TNG, allowing thus the coupling with HARPS--N. The main activity of the GIARPS project is therefore the modification of the configuration of GIANO that will be fed by a new train of preslit optics instead of the fibers. This will provide a significant improvement in the instrument performances, since the modal noise will be removed.
A rigid structure will fasten the GIANO dewar to the fork of the TNG (see Figure \ref{fig:structure}) by keeping the cryostat axis parallel to the elevation axis of the telescope. The structure has been thought in order to sustain a burden of about 2000 kg and it is rigid enough in order to not add vibration modes to those naturally generated by the movements of the telescope (Jitter, tracking etc.).

\begin{figure}
\centering
\resizebox{0.7\textwidth}{!}{
 \includegraphics{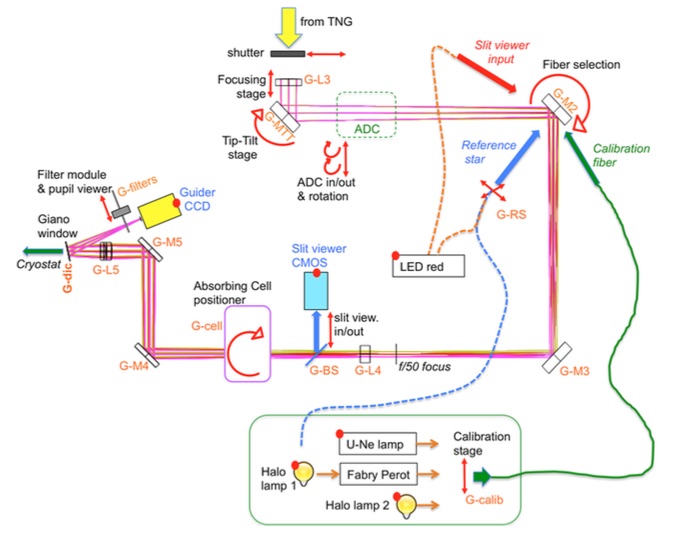}
}
\caption{Scheme of the new optics for GIANO, necessary to obtain the  and use it simultaneously with HARPS--N \cite{tozzietal2016}.}
\label{fig:preslit}       
\end{figure}

The new GIANO optics is fully described in \cite{tozzietal2016}: with reference to Figure \ref{fig:preslit} a brief description is given in the following.
The light coming from the Nasmyth focus of the TNG meets at first a dichroic that reflects the visible component toward the HARPS--N FEU and transmits the near IR component to GIANO. This dichroic is mounted on a slide in the Nasmyth box that allows to select the preferred observing mode. In the case of GIARPS observing mode, the dichroic is inserted. 
The near IR light is firstly directed towards the re-imaging module (the pick up module) inside the Nasmyth box of interface with the instruments. 
The pick up creates an intermediate focus below the de-rotator, in the volume previously allocated to SARG spectrograph \cite{grattonetal2002}. The light is then redirected towards the G-L3 module in order to correct and optimize the focusing of the stellar image onto the slit of the NIR spectrograph. Just after G-L3 there is a tip tilt mirror (G-MTT). The next optical elements, G-M2 (rotatory mirror) and G-M3 (fixed mirror) are used to select the calibration input from the calibration unit. The latter is beneath the preslit plane and is equipped with a U-Ne lamp, a halogen lamp and a Fabry Perot selectable by a translation stage. 
After that, mounted on a rotating stage, there will be inserted an absorption cell that will provide a stable reference spectrum for high precision radial velocity measurements. 
To minimize systematic errors the gas cell will be filled with gases (in particular acethylene, ammonia and an isotopologue of methane) at very low value of pressure aiming to reduce the pressure-induced line-shift and to exploit the intrinsically narrower lines. For this reason, for a given mixture of gases, a long cell filled at low pressure should be always preferred to a shorter cell filled at higher pressures, so we use the maximum space/length available within the volume previously allocated to SARG, in order to have a 1.5m-long cell. In this configuration the light is brought inside the GIANO dewar by means of a set of optics (the periscope G-M4/G-M5 plus the re-imaging lens G-L5) that allows also to focus the light onto the entrance slit of the spectrometer. Finally, a slit viewer allows in daytime to find the exact position of the slit on the guiding camera, providing a better centering of the star during the night-time observations.

\section{Radial velocities with GIARPS and Perspective}
As previously mentioned, the increasing interest in M-type stars is due to the fact that recent studies have determined that they are more likely to host rocky planets \cite{Beanetal2010}. In order to find habitable planets in orbit around solar-type stars, the RV technique has to achieve a precision of $10~cm/s$. This constraint is released searching around less massive stars, because the radial velocity (RV) amplitude of an Earth-mass planet around an M-dwarf is larger (about $1~m/s$). \\
While the solat-type stars are bright at wavelength shorter than 1~$\mu$m (visible region), the M-dwarfs are much fainter at optical wavelengths, because they have effective temperatures of 4000K or less and emit most of their energy at wavelength longer than 1~$\mu$m. So the near-infrared (NIR) instruments are crucial to observe these targets. The NIR region is also important for the RV technique, in fact the RV variation can be induced either by inhomogeneities on the stellar surface, like spots, or by the presence of a companion around the star. Acquiring NIR spectra allows to reduce the RV jitter related to the stellar activity of about 1/3 respect to the visible measurements, because the contrast between stellar spots or plagues and the rest of the stellar disk is reduced. Starting from this concept, a comparison between RV variations measured in the optical and NIR ranges can establish the origin of the RVs in an unambiguous way: if the variations are due to a companion, the signal amplitude has to be the same in every wavelength regions, while the variations due to the stellar activity is dependent on the wavelengths. A clear example is represented by the giant planet around the star TW Hya claimed by \cite{Setiawanetal2008}, later rectrated from CRIRES observations (\cite{Kaufletal2006}) that showed the decrease of the RV amplitude revealing the activity nature of the RV variations. \\
In this context GIARPS represents a very powerful instrument providing simultaneous spectra in optical (HARPS-N) and NIR (GIANO) regions. 
HARPS-N is equipped with two fibers, garanteeing an accurate localization of the wavelength in the detector with the simultaneous reference technique. During scientific observations the first fiber is fed with the star light, and from this spectrum the stellar RV is computed by referring to the wavelength solution determined at the beginning of the observing night. The second fiber is illuminated with the same spectral reference all the time, in order to measure possible instrumental drifts. \\
The RVs from GIANO spectra are measured by using the CCF (Cross Correlation Function) method \cite{carleoetal2016}, with the telluric spectrum as wavelength reference. The digital stellar and telluric masks are created from the acquired spectra, and then the RVs are computed for both stellar and telluric spectra by cross-correlating the masks with the spectra. The procedure also provides the bisectors of the CCFs and the bisector velocity spans as a first check of the origin of the RV variations. The use of this technique allows to reach a precision of about 10 $m/s$ for bright stars (H$\leq$5 mag) and about 70 $m/s$ for fainter stars (tipically H$\sim$9 mag). The introduction of the absoprion cell will allow to reach more precise RV measurements with internal errors of about 3$m/s$, due to the fact that absorption cell is more reliable in comparison with the instability of the telluric spectrum. 

\section{Conclusion}
Once GIARPS will work routinely at the telescope, TNG will have a high resolution spectroscopy station that will be unique in the northern hemisphere and up to the commissioning of NIRPS (the NIR counterpart of HARPS) at the 3.6m ESO Telescope, the unique in this world. The flexibility of the three observing modes of GIARPS: HARPS--N alone, GIANO alone and GIARPS itself will allow users to select the best wavelength range useful for their preferred science case. From small bodies of  the Solar System to the search for extrasolar planets will be the major science cases. For the latter, GIARPS will be the unique facility in this world that will allow to have simultaneously high precision radial velocity measurements in VIS (HARPS-N) and NIR (GIANO) wavelength range covering from 0.390 $\mu$m to 2.5 $\mu$m.

\end{document}